\documentclass[prb,nofootinbib,longbibliography,superscriptaddress,preprintnumbers,amsmath,amssymb,twocolumn,floats]{revtex4-2}
\usepackage{txfonts}
\usepackage{amssymb}
\usepackage{graphicx} 
\usepackage{sidecap}
\usepackage{CJK}
\usepackage{blindtext}
\usepackage{url}
\usepackage{epstopdf}
\usepackage{color}
\usepackage{mathtools}
\usepackage{amsmath,amssymb}
\usepackage{mathrsfs}
\usepackage{natbib}
\usepackage{bbm}
\usepackage{soul}

\usepackage{amsmath}
\numberwithin{equation}{section}
\begin{document}
\bibliographystyle{apsrev4-1}
\title{Observation of multiple surface states in naturally cleavable chiral crystal PdSbSe}

\author{Zhicheng Jiang}
\thanks{Equal contributions}
\affiliation{Shanghai Synchrotron Radiation Facility, Shanghai Advanced Research Institute, Chinese Academy of Sciences, Shanghai, 201210, China}
\affiliation{National Synchrotron Radiation Laboratory and School of Nuclear Science and Technology, University of Science and Technology of China, Hefei, 230026, China}

\author{Zhengtai Liu}
\thanks{Equal contributions}
\thanks{liuzt@sari.ac.cn}
\affiliation{Shanghai Synchrotron Radiation Facility, Shanghai Advanced Research Institute, Chinese Academy of Sciences, Shanghai, 201210, China}
\affiliation{National Key Laboratory of Materials for Integrated Circuits, Shanghai Institute of Microsystem and Information Technology (SIMIT), Chinese Academy of Sciences, Shanghai, 200050, China}

\author{Chenqiang Hua}
\thanks{Equal contributions}
\affiliation{Zhejiang Province Key Laboratory of Quantum Technology and Device, Department of Physics, Zhejiang University, Hangzhou 310027, China}
\affiliation{State Key Lab of Silicon Materials, School of Materials Science and Engineering, Zhejiang University, Hangzhou 310027, China}
\affiliation{Beihang Hangzhou Innovation Institute Yuhang, Xixi Octagon City, Yuhang District, Hangzhou 310023, China}

\author{Xiangqi Liu}
\thanks{Equal contributions}
\affiliation{School of Physical Science and Technology, ShanghaiTech University, Shanghai 200031, China}

\author{Yichen Yang}
\affiliation{State Key Laboratory of Functional Materials for Informatics, Shanghai Institute of Microsystem and Information Technology, Chinese Academy of Sciences, Shanghai 200050, China}

\author{Jianyang Ding}
\affiliation{State Key Laboratory of Functional Materials for Informatics, Shanghai Institute of Microsystem and Information Technology, Chinese Academy of Sciences, Shanghai 200050, China}

\author{Jiayu Liu}
\affiliation{State Key Laboratory of Functional Materials for Informatics, Shanghai Institute of Microsystem and Information Technology, Chinese Academy of Sciences, Shanghai 200050, China}

\author{Jishan Liu}
\affiliation{Shanghai Synchrotron Radiation Facility, Shanghai Advanced Research Institute, Chinese Academy of Sciences, Shanghai, 201210, China}
\affiliation{National Key Laboratory of Materials for Integrated Circuits, Shanghai Institute of Microsystem and Information Technology (SIMIT), Chinese Academy of Sciences, Shanghai, 200050, China}

\author{Mao Ye}
\affiliation{Shanghai Synchrotron Radiation Facility, Shanghai Advanced Research Institute, Chinese Academy of Sciences, Shanghai, 201210, China}
\affiliation{National Key Laboratory of Materials for Integrated Circuits, Shanghai Institute of Microsystem and Information Technology (SIMIT), Chinese Academy of Sciences, Shanghai, 200050, China}

\author{Ji Dai}
\affiliation{ALBA Synchrotron Light Source, Barcelona, 08290 Spain}

\author{Massimo Tallarida}
\affiliation{ALBA Synchrotron Light Source, Barcelona, 08290 Spain}

\author{Yanfeng Guo}
\email{guoyf@shanghaitech.edu.cn}
\affiliation{School of Physical Science and Technology, ShanghaiTech University, Shanghai 200031, China}

\author{Yunhao Lu}
\email{luyh@zju.edu.cn}
\affiliation{Zhejiang Province Key Laboratory of Quantum Technology and Device, Department of Physics, Zhejiang University, Hangzhou 310027, China}
\affiliation{State Key Lab of Silicon Materials, School of Materials Science and Engineering, Zhejiang University, Hangzhou 310027, China}

\author{Dawei Shen}
\email{dwshen@ustc.edu.cn}
\affiliation{National Synchrotron Radiation Laboratory and School of Nuclear Science and Technology, University of Science and Technology of China, Hefei, 230026, China}

\begin{abstract}
%High-order fermions and long Fermi arcs have been theoretically and experimentally reported in chiral crystals with the $P2_13$ space group, such as CoSi, PtGa and AlPt, attracting significant interest. Compared to these binary chiral compounds, ternary chiral compound in space group $P2_13$ exhibit larger and more complex crystal frameworks, while their chiarl topological properties remain unclear. The primary challenge lies in the more complex orbitals and much smaller periods in momentum space.
Chiral multifold fermions in solids exhibit unique band structures and topological properties, making them ideal for exploring fundamental physical phenomena related to nontrivial topology, chirality, and symmetry breaking. However, the challenge of obtaining clean, flat surfaces through cleavage has hindered the investigation of their unique electronic states. In this study, we utilize high-resolution angle-resolved photoemission spectroscopy and density functional theory calculations to investigate the low-energy electronic structure of the cleavable single-crystal PdSbSe. Our combined experimental and theoretical analysis reveals the presence of multifold degenerate fermions within this chiral crystal. We also observe multiple chiral Fermi arc surface states and spin-splitting behavior in the associated bulk bands. These findings provide unique insights into chiral, multifold fermionic states in easily cleavable crystals and offer a robust platform for further research into their unique electronic properties and potential applications in novel electronic devices.
\end{abstract}

\maketitle
\clearpage
%---------------------------introduction-------------------------
\section{Introduction}

Chirality, a geometric property of asymmetrical objects that are not superimposable on their mirror images, is a fundamental concept in modern science. Meanwhile, the chirality inherent in crystal structures can profoundly influence the electronic states of materials, thereby impacting their quantum behaviors. For instance, the chirality in topological semimetals gives rise to various fascinating quantum phenomena such as chiral edge states~\cite{mancini2015observation, susstrunk2015observation,young2014tunable}, Weyl fermions, chiral Majorana fermion~\cite{liu2018robust,jiao2020chiral,tanaka2009manipulation}, and chiral symmetry-protected states~\cite{mancini2015observation}. Among these, Weyl fermions in topological crystal materials exhibit conspicuous chirality-related features, where the breaking of time-reversal-symmetry (TRS) results in pairs of Weyl points with opposite chirality~\cite{li2016chiral,parameswaran2014probing,ma2017direct,lv2015observation,lv2015experimental}, characterized by signed topological invariant Chern numbers $\pm{1}$. Additionally, higher-fold chiral fermions, such as three-, six- and eight-component fermions are suggested to feature larger Chern numbers than those observed in Weyl semimetals~\cite{chiu2016classification,liu2014discovery,lv2019observation}. 

\begin{figure*}[htb]
\centering
\includegraphics[width=14cm]{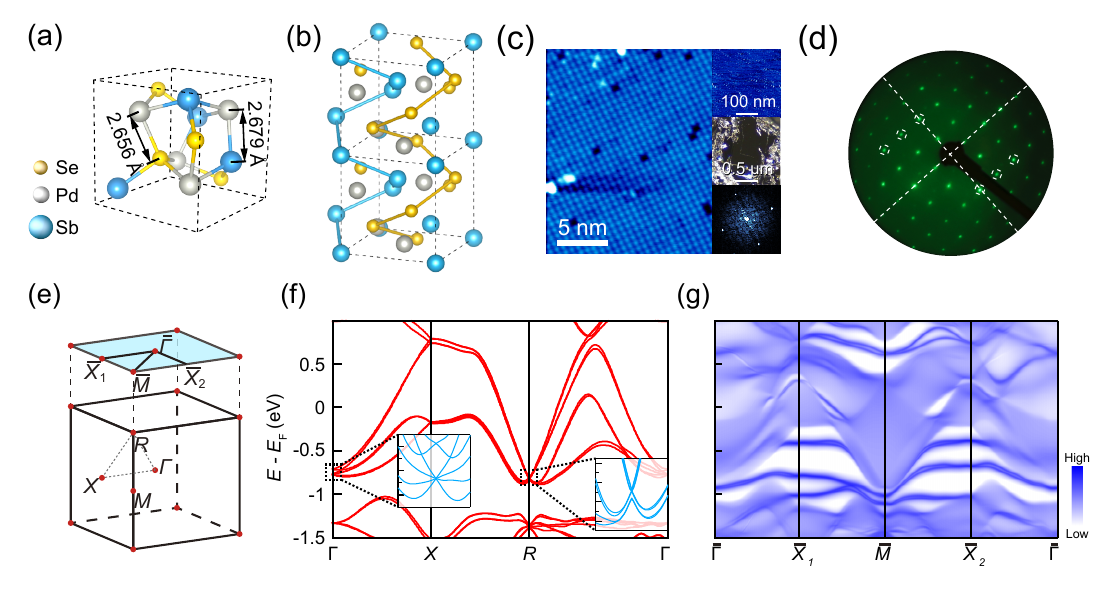}
\caption{
Crystal structure and calculated eletronic electronic structure of PdSbSe. 
(a) PdSbSe crystal structure within the unit cell. Gold, gray and blue spheres represent the Se, Pd, and Sb atoms, respectively. 
(b) Chiral crystal arrangement along the (001) direction. 
(c) High-resolution STM image (20 $\times$ 20 nm$^2$) showing the atomic arrangement, with an inset large-scale STM image (300 $\times$ 300 nm$^2$), a photograph of the (001) cleave surface, and the corresponding FFT image. 
(d) LEED pattern of cleaved (001) surface measured with electron energy 300 eV. 
(e) Three-dimensional Brillouin zone illustration of PdSbSe. 
(f) \textit{Ab-initio} calculated bulk band structure along high-symmetry lines with spin-orbit coupling (SOC). The inset shows an enlarged view of the band structure around $\Gamma$ and $R$ points. 
(g) Calculated band structure with surface states.
}
\label{fig:Figure 1}
\end{figure*}

Recently, the chiral semimetals family in space group 198 (namely P2$_1$3) has been the subject of extensive study owing to its preservation of higher-fold degenerate band nodes~\cite{takane2019observation,rao2019observation,rees2020helicity,chang2017unconventional,schroter2019chiral,yao2020observation,ma2021observation,sessi2020handedness,lv2019observation,schroter2020observation,cochran2023visualizing}. Similar to Weyl fermions, these exotic fermions at degenerate points can transport between points with opposite Chern numbers via channels known as topological surface Fermi arcs. Meanwhile, these surface states are directly connected to the topological properties at points of high-fold degeneracy, where the number of Fermi arcs extending from each node exactly corresponds to the magnitude of the Chern number~\cite{bradlyn2016,tang2017multiple}. From an applied science perspective, chiral multifold degenerate fermion materials possess unique electronic properties such as high electron mobility and the quantized circular photogalvanic effect, both of which hold tremendous potential for the development of novel electronic devices~\cite{de2017quantized,chang2017unconventional,ni2021giant,rees2020helicity}. However, to date, despite several photoemission spectroscopic studies on chiral crystals, definite identification of chiral multiple-fold nodes and Fermi arcs remain scarce~\cite{cochran2023visualizing}. For most chiral crystals, the challenge often stems from the inability to obtain clean and flat surfaces through cleavage. Poor surface conditions typically obscure the clear identification of band dispersion and Fermi-arc surface states in spectroscopic analyses~\cite{tang2020atomically}.

In this paper, we designed and synthesized a chiral topological semimetal candidate, PdSbSe, which exhibits substantial spin-orbit interaction and can be readily prepared with clean and flat surfaces by cleavage in ultrahigh vacuum. Using high-resolution angle-resolved photoemission spectroscopy (ARPES), we identified both distinct multifold degenerated band crossings with high-order Chern numbers and Fermi-arc surface states. Additionally, we also discovered multiple types of surface states and complex hybridized behavior involving chiral Fermi arcs in this compound. These findings not only enhance our comprehension of novel fermionic systems, but also provide an important platform and reference for the development of new quantum materials.

\begin{figure}[t]
\centering
\includegraphics[width=8.3cm]{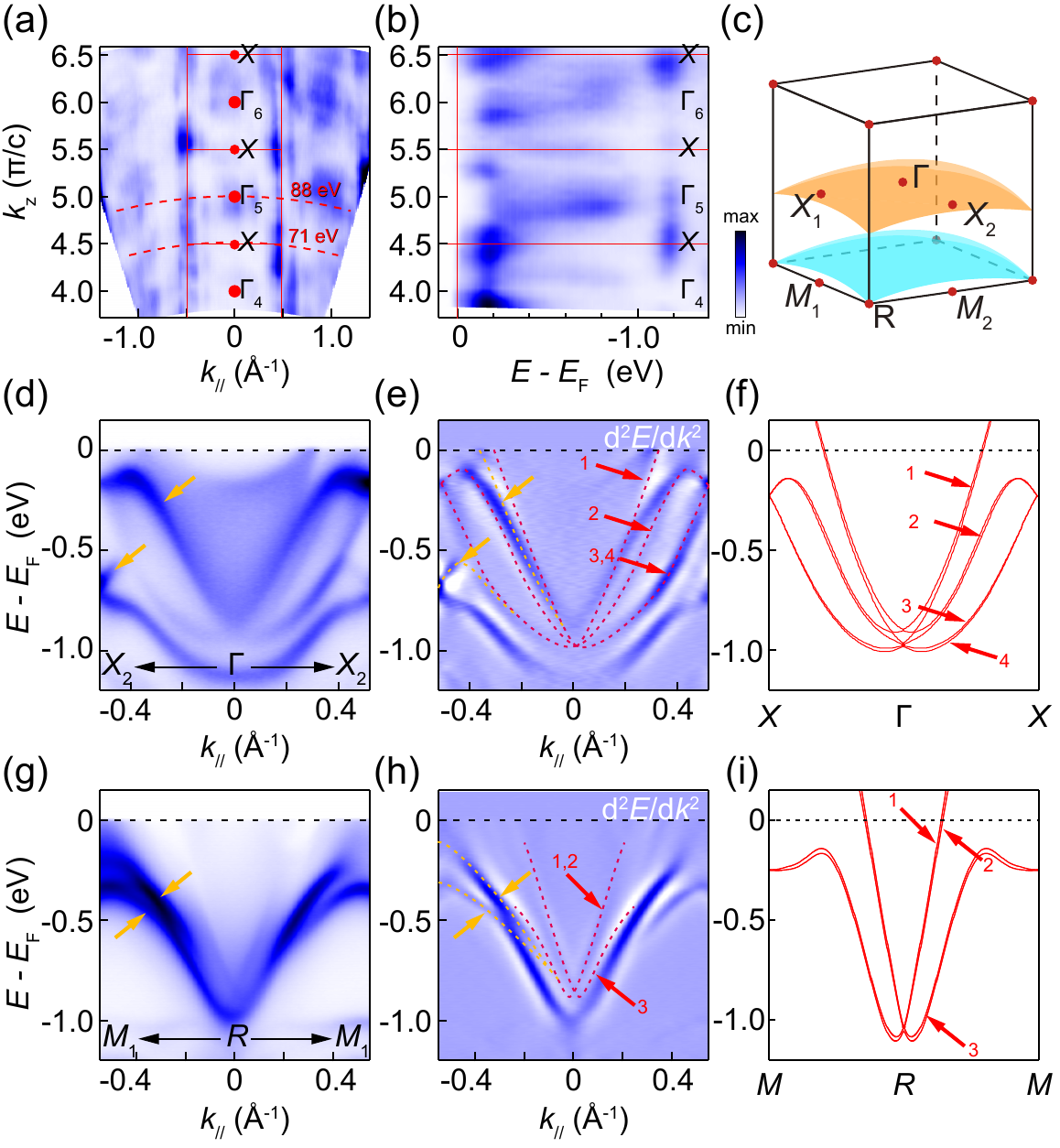}
\caption{
The photon-energy-dependent ARPES data. 
(a) Experimental constant energy contour at $E_B$ = 0.9 eV measured by photon energy $h\nu$ = 60 $\sim$ 160 eV. Periodic Brillouin zones and photon energies corresponding to high-symmetry cuts are appended on this map. The inner potential is set to $V_0$ = 5~eV.
(b) $E-k_z$ ARPES band dispersion extracted at $k_{//}$=0 \AA$^{-1}$, clearly showing the periodic dispersion pattern of the electron pockets at bulk $\Gamma$ points. 
(c) Model of unit Brillouin zone, with the colored curved surface indicating the momentum regions detected by 88 and 71~eV incident photons. 
(d) ARPES intensity plot of band dispersion measured along $X_2$-$\Gamma$-$X_2$ direction using 88~eV photons.
(e) Second-derivation plot of (d), with bulk bands denoted by red dashed lines and arrows. Extra surface states are marked by yellow arrows in (d-e).
(f) Bulk band calculation along $X$-$\Gamma$-$X$ direction.
(g)-(i) Same configuration along the $M_2$-$R$-$M_2$ direction shown in (d)-(f).
(a)-(b),(d)-(e) and (g)-(h) share the same colorbar.
}
\label{fig:Figure 2}
\end{figure}

\section{Method}
High-quality PdSbSe single crystals were synthesized by mixing high-purity substances including palladium (99.9$\%$), antimony (99.999$\%$) and selenium (99.999$\%$) in a molar ratio of Pd:Sb:Se=1:1:1. The mixture was sealed in a quartz tube, heated to 800 $^{\circ}$C in a furnace for 30 hours, and then slowly cooled to 650 $^{\circ}$C at a rate of 2 K/h. The composition of crystals was examined through energy-dispersive x-ray (EDX) spectroscopy. Phase and crystalline quality examinations were performed on a single-crystal x-ray diffractometer equipped with a Mo K$\alpha$ radioactive source ($\lambda$ = 0.71073 \AA). ARPES measurements were performed at the 03U beamline of Shanghai Synchrotron Radiation Facility (SSRF)~\cite{yang2021high,sun2020performance} and LOREA beamline of ALBA synchrotron. At SSRF BL03U, measurements utilized a Scienta Omicron DA30 hemispherical analyzer with energy resolution of 10$\sim$15 meV (photon energy-dependent) and angular resolution of 0.2$^\circ$, using linearly horizontal  and vertical light. Samples were cleaved \textit{in situ} along high-symmetry crystalline planes and measured under ultrahigh vacuum ($\textless$8$\times$10$^{-11}$ Torr) at a stabilized temperature of 15 K. Parallel experiments at ALBA LOREA employed an MB Scientific A-1 momentum-resolving spectrometer with 0.2$^\circ$ angular resolution, offering both linear (horizontal/vertical) and circularly (right-/left-handed) polarized light. These measurements were conducted under $\textless$5$\times$10$^{-10}$ Torr vacuum at 20 K.First-principles calculations were performed in the Vienna $Ab$ $initio$ Simulation Package (VASP) by implementing projector-augmented wave (PAW) pseudopotential with exchange-correlation functional of generalized gradient approximation of Perdew, Burke, and Ernzerhof ~\cite{kresse1996efficiency,chen2018layer,kresse1999ultrasoft,perdew1996generalized}. Besides, SOC was taken into consideration as a perturbation term~\cite{SOC1977} in the band structure calculation of PdSbSe. The convergence criteria for energy and force were set to 10$^{-6}$ eV and 0.01 eV/Å, respectively. The plane-wave cutoff energy was set at about 400 eV and 15 $\times$ 15 $\times$ 15 $\Gamma$-centered $k$-mesh of BZ was adopted. Maximally localized Waniner functions~\cite{mostofi2014updated} based on Sb $p$ and Pd $d$ orbitals were generated using WANNIERTOOL~\cite{wu2018wanniertools} to obtain the surface state spectrum.

\section{Results and Discussion}
PdSbSe is a topological semimetal that crystalizes in the cubic space group $P2_13$ (No. 198) with a lattice constant of $a$ = 6.565 \AA. Figure~1(a) illustrates the unit cell of PdSbSe, and Fig.~1(b) highlights its chiral crystal structure. During the ARPES measurement, we cleaved the PdSbSe crystal along the (001) direction, breaking the Pd-Se and Pd-Sb chemical bonds, leaving a square-like Pd atom surface, as shown by the scanning tunneling microscopy (STM) image in Fig.~1(c). The middle inset of Fig.~1(c) presents a photograph of a typical PdSbSe (001) cleaving surface. Although crystals with $P2_13$ space group are generally difficult to cleave, we were able to obtain a high-quality pristine cleaving plane because the Pd-Sb and Pd-Se bonds are relatively longer than those in binary chiral crystals (PtGa, PdGa, etc.). The low energy electron diffraction (LEED) image acquired on the (001) cleaving surface [Fig. 1(d)] reveals the chiral distribution of bright spots, similar to previous reports in binary chiral crystals~\cite{schroter2020observation}. The cubic Brillouin zone (BZ) and projected (001) plane are displayed in Fig.~1(e). Protected by helical symmetry $\mathcal{C}_{nd}$ (where $n$ and $d$ represent the rotation number and axis, respectively), combined with time-reversal symmetry $\mathcal{T}$, point-like bands degeneracies emerge at high symmetry momentum points. For the space group $P2_13$, the degeneracy points are located at $\Gamma$ and $R$ points, as highlighted in Fig.~1(f). These points act as sources or sinks of quantized Berry flux, resulting in positive or negative integral Chern numbers, with corresponding numbers of topological surface states. Considering spin-orbit coupling (SOC), the Chern number doubles, reaching a maximum of $\pm$4 in PdSbSe. First-principles calculations of bulk bands involving SOC reveal four-fold and six-fold bands degeneracies at the $\Gamma$ and $R$ points, respectively. However, the band splitting is too small to be distinguished, and we can only identify two distinct bands around $R$ point, as shown in the zoomed-in inset of Fig.~1(f). Figure~1(g) shows the calculated band structure with surface states on the (001) projected surface, where several topological surface states are expected to appear. 

We began by focusing on the degeneracy points in the bulk band of PdSbSe, using ARPES measurements with photon energies in the range 60-160~eV. The resulting $k_{//}$-$k_z$ map at binding energy ($E_B$) of 0.9~eV is shown in Fig.~2(a). Despite the suppression of periodic features owing to chiral symmetry, high-symmetry points remain discernible, as highlighted by solid frames and dots. To better visualize the periodic dispersion, we extract the $E$-$k_z$ spectra along $k_{//}$=0, which is presented in Fig.~2(b). Here, the pronounced electron pockets help to identify the $\Gamma$ and $X$ points. Additionally, we compared the $k_z$ maps along $X_1$-$\Gamma$-$X_1$ and $X_2$-$\Gamma$-$X_2$ directions in Fig.~S1 (see the Supplemental Material, SM~\cite{SI}), which helps identify the photon energies corresponding to the high-symmetry planes. The degeneracy bands at the $\Gamma$ point can be detected by 88~eV photons, while the $R$ points correspond to 71 eV, as illustrated in Fig.~2(c). Because of the anisotropic surface states induced by the chiral structure, which mix with the bulk bands, we use different subscripts (e.g. ``$X_1$'' and ``$M_1$'') to label these high symmetry points in Fig.~2(c) for clarity. Figures~2(d)-(f) provide a direct comparison of the ARPES spectra, second-derivative spectra, and bulk band calculations along the $X_2$-$\Gamma$-$X_2$ direction. Three bulk bands, indicated by red arrows in the second-derivative spectra [Fig.~2(e)], align closely with the calculations [Fig.~2(f)], where a fourfold degeneracy point is expected according to our calculation. However, since the energy splitting between band 3 and 4 is too small to be resolved, we performed a detailed analysis of the energy distribution curves (EDCs) across the largest band splitting between bands 3 and 4, where two peaks can be resolved, as shown in Figs.~S2(a)-S2(d) (see the SM~\cite{SI}). Similarly, the multiband crossings are evident along $M_1$-$R$-$M_1$, as shown in Figs.~2(g)-(i). The bulk degenerate bands, marked by dashed red lines in Fig.~2(h), show very minor band splittings, consistent with the calculations. These bulk band splittings are difficult to resolve from either the EDCs or momentum distribution curves (MDCs), as shown in Figs.~S2(f)-S2(h) (see the SM~\cite{SI}). However, a very small band splitting between band 1 and 2 can be resolved from the dispersion along $R$-$M_2$ direction, as shown in Fig.~S2(i)(see~\cite{SI}. Consequently, we can only observe the characteristic indicative of a triple degenerate point, rather than the expected six-fold degenerate point. Remarkably, in the spectra shown in Fig.~2(d) and 2(g), additional bands with stronger spectral weight, marked by yellow arrows, are visible below the energy levels of multiple degenerate points. These bands significantly obscure the visibility of the bulk degenerate bands. A comparative analysis of the calculation results across different $k_z$ positions reveals that the bright and sharp bands are not caused by $k_z$ band broadening. The dispersionless feature observed in the $k_z$ map [Fig.~2(a)] and the nonmirror-symmetric feature in the $k_x$-$k_y$ map (Fig.~3) indicate that these additional bands are primarily associated with surface states. More comparison of bulk bands along different high-symmetric directions can be found in Fig.~S3 within the SM~\cite{SI}.

\begin{figure*}[htb]
\centering
\includegraphics[width=17cm]{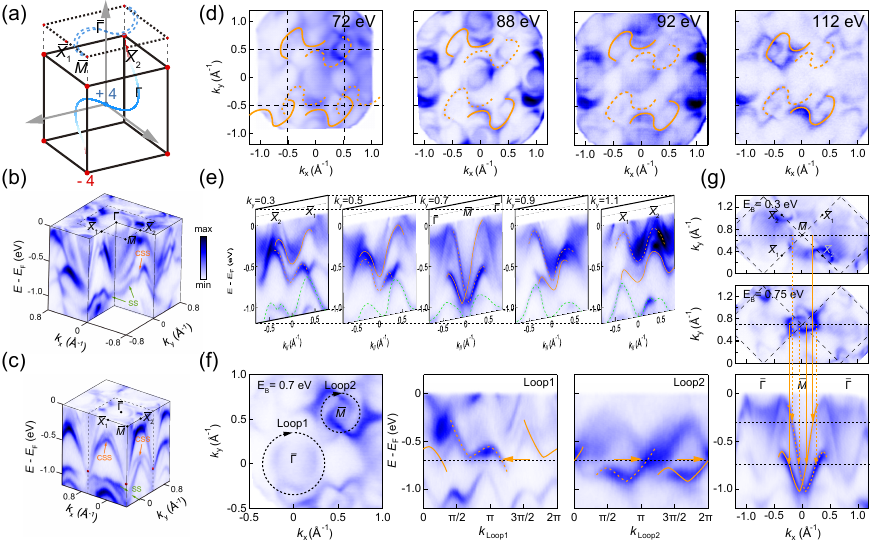}
\caption{
(a) Schematic diagram showing the distribution of nonzero topological charges at degeneracy points and the corresponding chiral surface states. 
(b)-(c) 3D intensity plot of the band dispersion around $\overline{\Gamma}$ and $\overline{M}$ points, with topological chiral surface states (CSS) marked by yellow arrows and trivial surface states (SS) marked by green arrows.
(d) Experimental constant energy contours of (001) cleave surface taken at $E_B$ = 0.63 eV, measured with photon energies of 72, 88, 92 and 112 eV. The appended yellow curves denote the chiral Fermi arcs surface states. 
(e) ARPES intensity plots of band dispersion along the diagonal direction of the Brillouin zone parallel to $\overline{\Gamma}$-$\overline{M}$-$\overline{\Gamma}$. 
(f) Enlarged view of the experimental constant energy contour at $E_B$ = 0.7~eV. Dashed lines with arrows represent two clockwise circular paths loop1 and loop2 around the $\Gamma$ and $M$ points.
ARPES intensity plot of band dispersion along loop1 and loop2 are also shown side-by-side, with yellow arrows indicating the CSSs. 
(g) Constant energy contours at $E_B$ = 0.3 eV (top) and 0.75 eV (middle). The bottom panel shows the ARPES measured band dispersion along the diagonal direction of Brillouin zone ($\overline{\Gamma}$-$\overline{M}$-$\overline{\Gamma}$). Yellow lines and arrows mark the position of CSSs.
(b)-(g) use a common colorbar.
}
\label{fig:Figure 3}
\end{figure*}

To confirm the distribution of the additional dispersion in PdSbSe, characterized by strong spectral weight, we present the three-dimensional plot of the photoemission spectra in Figs.~3(b) and 3(c). The chiral surface states (CSS) near the $\overline{\Gamma}$ and $\overline{X}$ merge deeply into the shadow of the projected bulk bands, similar to the phenomenon observed in PdBiSe~\cite{lv2019observation}, where the chiral surface states are completely obscured by bulk bands projection from $k_z$. Nevertheless, these CSSs are distinguishable away from the degeneracy points in PdSbSe, as indicated by yellow arrows in Fig.~3(b) and 3(c), showing their location at the outer edge of bulk bands. In Fig.~3(d), we present constant energy contours at $E_B$ = 0.63 eV, taken with different incident photon energies $h\nu$ = 72 $\sim$ 112 eV. The ``S''-like features maintain a consistent shape and position across all photon energies, except for variations in relative intensity. As denoted by yellow solid and dash lines (representing the extension of the CSSs to different Brillouin zones), two clockwise ``S''-shaped arms encircle the electron pocket at the $\overline{M}$ points and finally sink near the $\overline{\Gamma}$ point in diagonal direction. Further, in Fig.~3(g), we illustrate the variations of these surface states around the $\overline{M}$ point for two different binding energies. Our ARPES measurements clearly depict the trend of the surface states, and the corresponding surface band dispersion along $\overline{\Gamma}$-$\overline{M}$ direction is distinguishable. Figure~3(e) shows that these surface states are distributed asymmetrically in the positive and negative \textit{k}-regions, deviating from the diagonal direction where $k_y$ = 0.7 \AA$^{-1}$ intersects the $\overline{M}$ point. This distribution exhibits near-mirror symmetry in the $k$- and $k$+ regions on the opposite side, highlighting the chiral nature of these Fermi arcs. We further compare the experiments under different polarizations, including linear horizontal (LH), linear vertical (LV), right-hand circular (C+) and left-hand circular (C-) polarizations {as shown in Fig. S4(a)-S4(d) with the SM~\cite{SI}}, to exclude the effect of matrix elements. While the intensity distributions changed significantly, the chiral nature remains unchanged under different light polarizations. Additionally, we present similar band dispersion with Fig.~3(e) under LV polarization in Fig.S4(e)-S4(f) (see the SM~\cite{SI}), where asymmetrical features are also observed on the both side of $\overline{\Gamma}$-$\overline{M}$-$\overline{\Gamma}$, further confirming the chiral nature of CSS in PdSbSe. Moreover, we display the dispersions along clockwise closed loops around $\overline{\Gamma}$ and $\overline{M}$ points in Fig.~3(f). The surface bands around these two points show opposite trends, aligning with the calculated opposite Chern numbers. Additionally, we identify another low-lying surface band, denoted by green-dashed lines, which is located beneath these chiral surface states and shows a significant correlation with them. Note that additional surface states, not connect to the topological nontrivial points, are attributed to trivial surface states (SS), marked by green arrows and dash lines.

\begin{figure*}[htb]
\centering
\includegraphics[width=14cm]{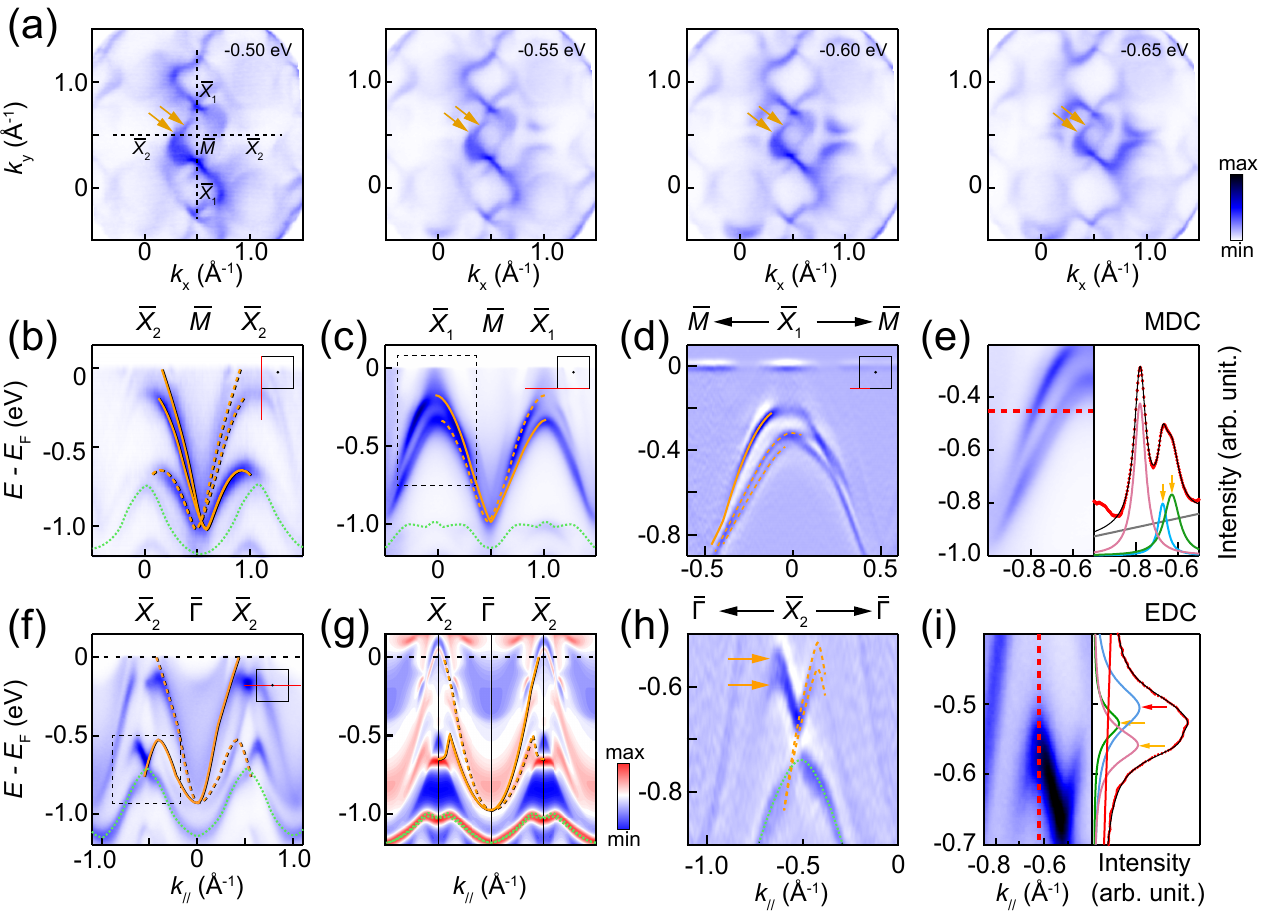}
\caption{
(a) ARPES constant energy contours around $\overline{M}$ point taken at $E_B$= 0.5 to 0.65 eV by step size of 0.05 eV, with CSSs marked by yellow arrows. 
(b)-(c) ARPES intensity plot of band dispersion along (b)$\overline{X_2}$-$\overline{M}$-$\overline{X_2}$ and (c)$\overline{X_1}$-$\overline{M}$-$\overline{X_1}$ directions. 
(d) Second-derivative plot of the enlarged region [marked by dash frame in (c)] around the $\overline{X_1}$ point.
(e) MDC analysis of CSSs around the $\overline{X_1}$ point, dashed-red line in the left half panel represents the MDC position in the right panel. Colored curves are Lorentz peaks used to fit the MDC. Fitted curve are marked by the dark lines.
(f) Experimental $E$-$k$ spectrum along $\overline{X}_2$-$\overline{\Gamma}$-$\overline{X}_2$ direction.
(g) Calculated band structure with surface states along the same direction as in (f). Red in the colorbar represent the max contribution of surface states.
(h) Second derivative spectrum of enlarged region [marked by dash frame in (f)] around $\overline{X}_2$ point.
(i) EDC analysis of CSSs around $\overline{X_2}$ point, dashed-red line in the left half panel represents the EDC position in the right panel. Colored curves are the Lorentz peaks used to fit the EDC. Fitted curve are marked by dark lines.
All figures except (g) share the same colorbar.
}
\label{fig:Figure 4}
\end{figure*}

The calculation indicates a Chern number of 4 around $\overline{\Gamma}$ and $\overline{M}$, suggesting that four CSSs spiral from these points. However, only two CSSs were identified in the analysis above. This discrepancy arises because the splitting of the surface states is too small to be resolved. Through a detailed analysis of the constant energy maps at different binding energies, we observed a relatively larger splitting at $E_B = 0.5~\mathrm{eV}$ around the $\overline{M}$ point, as shown in Fig.~4(a), where the CSSs split into two bands. Further analysis of the cuts through this split band, illustrated in Fig.~4(b), reveals that the CSSs gradually split into two bands as they approach the Fermi level. In Figs.~S4(c)-S4(e) within the SM~\cite{SI}, we analyzed the band structure along the $\overline{M}_1$-$\overline{X}$-$\overline{M}_1$ direction. Although the CSS splitting is subtle, it can still be discerned using the second-derivative spectrum, as presented in Fig.~4(d). In Fig.~4(e), MDC analysis and peak fitting of the surface states near $k = 0$ clearly reveal two split peaks. The density functional theory (DFT) calculations along this direction, shown in Fig.~S6 with the SM~\cite{SI} also indicates a similar CSS splitting. While there are slight deviations between the theoretical and experimental positions, the overall trends are consistent (see the SM~\cite{SI}). This deviation may originate from the cleaved surface. In the Fig.~S6 within the SM~\cite{SI}, we compare the effects of different cleavage surfaces on the band structure. We found that only the Pd-terminated surface (denoted as "cut-edge 1" in Fig.~S6) could adequately explain the formation of the additional SS band marked in green in Figs.~3 and 4, which is not connected to the topologically protected multi-degenerate points. Additionally, surface breaking chemical bonds, surface relaxation, and surface charges may also contribute to the discrepancies between the calculations and experiments. Furthermore, we compared the band structures along the $\Gamma$-$X_2$ direction. The CSSs along this direction correspond well to the topological surface states calculated in Fig.~4(g), though there are still slight deviations in the SS positions, as shown in Figs.~4(f)-4(g). We also analyzed the splitting of the CSSs along this direction. Owing to the proximity of the CSSs to bulk states, the splitting of the surface states is not prominent. Even in the second-derivative spectrum in Fig.~4(h), it remains challenging to distinguish. By performing peak fitting of the EDC along this surface state, we identified three Lorentzian peaks. Among these, the highest-energy peak originates from the bulk state near the CSSs, while the remaining two peaks correspond to the split surface states.
These features confirm the Chern number of $\Gamma$ and $R$ point, reaching a maximum magnitude of $\lvert$C$\rvert$ = 4 in PdSbSe. Notably, the two CSSs around the $\overline{X_2}$ point depicted in Fig.~4(f) and 4(h) extend to lower energy and overlap with SSs. This differs from the observation in PdBiSe~\cite{lv2019observation}, in which the chiral surface states merge into the bulk band projection. Additionally, we observed surface band hybridization at the crossing points, distinguishable at various photon energies. These findings indicate complex interactions between different surface states, which could significantly contribute to carrier transport across varied surface states.

%---------------------------summary---------------------------------
\section{Conclusions}
In conclusion, we have identified a unique type of cleavable chiral crystal. High-resolution ARPES measurements allowed us to experimentally demonstrate its multifold degenerate points at $\Gamma$ and $R$. Additionally, we confirm the presence of chiral Fermi arcs connecting the $\overline{\Gamma}$ and $\overline{M}$ points within this system. Further analysis of the chiral orientation enabled us to verify a maximum Chern number of $\lvert$C$\rvert$ = 4, with opposite signs at the $\Gamma$ and $R$ points. Moreover, we unveil complex hybridization among various types of surface states, which could potentially contribute to novel, exotic transport phenomena. These findings not only provide insights into chiral surface states, but also establish a robust platform for further investigations into multifold fermions.

\begin{center}
\textbf{ACKNOWLEDGMENTS}
\end{center}

This work was supported by the National Science Foundation of China (Grant Nos. U2032208, 11874264,92065201), National Key R\&D Program of China (Grants No. 2023YFA1406304, 2019YFE0112000), the Natural Science Foundation of Shanghai (Grant No. 14ZR1447600), the Zhejiang Provincial Natural Science Foundation of China (LR21A040001, LDT23F04014F01), the Shanghai Science and Technology Innovation Action Plan (Grant No. 21JC1402000), and the China National Postdoctoral Program for Innovative Talents (BX20240348). Y. F. Guo acknowledges the starting grant of ShanghaiTech University and the Program for Professor of Special Appointment (Shanghai Eastern Scholar). Part of this research used Beamline 03U of the Shanghai Synchrotron Radiation Facility, which is supported by ME$^2$ project under contract No. 11227902 from National Natural Science Foundation of China. The authors also thank the support from Analytical Instrumentation Center (\#SPST-AIC10112914). Part of experiments were performed at LOREA beamline at ALBA Synchrotron with the collaboration of ALBA staff. LOREA beamline is co-funded by the European Regional Development Fund (ERDF) within the ``Framework of the Smart Growth Operative Programme 2014-2020”.

%\begin{center}
%\textbf{ACKNOWLEDGMENTS}
%\end{center}

\begin{center}
\textbf{Data Availability}
\end{center}

The data are available from the authors upon reasonable request.

\bibliographystyle{apsrev4-2}
%\bibliography{ref}
%

%--------------------------------------------------------------------------
%                        Supplement Material
%--------------------------------------------------------------------------
\clearpage
\onecolumngrid
\begin{center}
    \textbf{\large Supplementary material for ``Observation of multiple surface states in naturally cleavable chiral crystal PdSbSe''}\\[.2cm]
\end{center}
%---------------------------------------------------------------------------
\maketitle
\setcounter{equation}{0}
\setcounter{section}{0}
\setcounter{figure}{0}
\setcounter{table}{0}
\setcounter{page}{1}
\renewcommand{\theequation}{S\arabic{equation}}
\renewcommand{\thesection}{ \Roman{section}}

\renewcommand{\thefigure}{S\arabic{figure}}
\renewcommand{\thetable}{\arabic{table}}
\renewcommand{\tablename}{Supplementary Table}

\renewcommand{\bibnumfmt}[1]{[S#1]}
\renewcommand{\citenumfont}[1]{#1}
\makeatletter

\maketitle

\setcounter{equation}{0}
\setcounter{section}{0}
\setcounter{figure}{0}
\setcounter{table}{0}
\setcounter{page}{1}
\renewcommand{\theequation}{S-\arabic{equation}}
\renewcommand{\thesection}{ \Roman{section}}

\renewcommand{\thefigure}{S\arabic{figure}}
\renewcommand{\thetable}{\arabic{table}}
\renewcommand{\tablename}{Supplementary Table}

\renewcommand{\bibnumfmt}[1]{[S#1]}
\makeatletter

\maketitle

\section{$k_z$ dependent measurements}
Comparison of $k_z$ maps along different directions is shown in Fig. S1, we perform the $k_z$ maps along the $\overline{\Gamma}$-$\overline{X_1}$ and $\overline{\Gamma}$-$\overline{X_2}$ directions (Fig. S1 (a, b)), focusing specifically on the energy level where $E_B$ = 0.73 eV (dash lines in Fig. S1(c, d)). This energy level is notable for its minimal surface state density, which enhances the clarity in distinguishing bulk characteristics. Our analysis has revealed a more pronounced dispersion along these directions, particularly at the $\overline{\Gamma}$ position, which is clearly discernible. Additionally, these $k_z$ features can be effectively differentiated within the second Brillouin zones.

%\begin{center}
\begin{figure*}[htb]
\centering
\includegraphics[width=14cm]{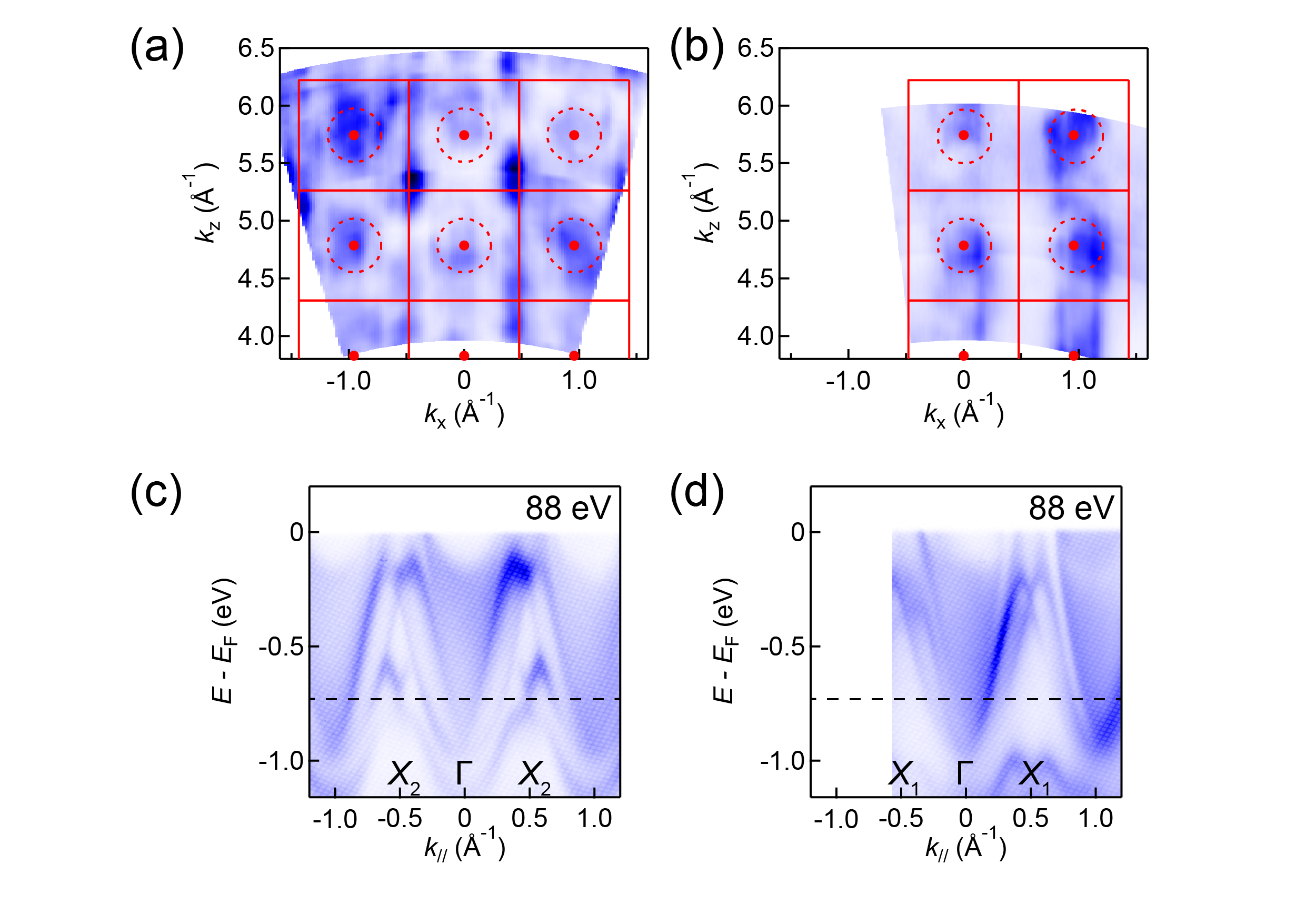}
\caption{
(a-b) ARPES intensity plot of $k_{//}$-$k_z$ map along (a) $\overline{X_2}$-$\Gamma$-$\overline{X_2}$ and (b) $\overline{X_1}$-$\Gamma$-$\overline{X_1}$ directions measured at $E_B$ = 0.73 eV;
(c-d) ARPES intensity plots of band dispersion along (c) $\overline{X_2}$-$\Gamma$-$\overline{X_2}$ and (d) $\overline{X_1}$-$\Gamma$-$\overline{X_1}$ directions, dark dash lines represent the energy $k_//$-$k_z$ maps are taken.
}
\label{fig:Figure S1}
\end{figure*}

\section{Bulk bands analysis}
In the $X_2$-$\Gamma$-$X_2$ direction, we can clearly distinguish three bands without considering the SOC splitting, as shown in Fig. S2(a). We further conducted a detailed band analysis and MDC analysis for the region of $E_B$=0.2~0.4 eV, allowing us to separate the bulk bands from the surface states (Fig. S2(c)). The red triangular arrows indicate the bulk bands, while the yellow arrows denote the surface states. However, bands 3 and 4 remain degenerate in these spectra. To effectively distinguish between bands 3 and 4, we performed an EDC analysis (Fig. S2(d)) along the direction indicated by the red solid line in Fig. S2(b). We observed that at $E_B$=0.9 eV, the band splits into two distinct peaks, which correspond to the calculated splitting between bands 3 and 4. Consequently, the quadruple degeneracy at the $\Gamma$ point can be resolved.
Additionally, we analyzed the bands along the $M_1$-$R$-$M_1$ direction. However, the bulk bands in this direction are not particularly clear, and the high-intensity surface states obscure the band 3, making it difficult to distinguish the bulk bands effectively. From Fig. S2(h), we can more clearly identify the bulk band labeled as 3, but the degenerate bulk bands labeled as 1 and 2 cannot be well distinguished. In contrast, the spectra along the $R$-$M_2$ direction allow for better separation of surface and bulk bands, as shown in Fig. S2(i). The figure displays the MDC curves between $E_B$=0.2~0.4 eV, where the bulk bands labeled as 1, 2, and 3, marked with red triangles, are well separated, indicating at least a triple degeneracy at this point. The splitting introduced by SOC near this point is too small to be resolved even computationally, so we can only distinguish three bulk bands.

\begin{figure*}[htb]
\centering
\includegraphics[width=14cm]{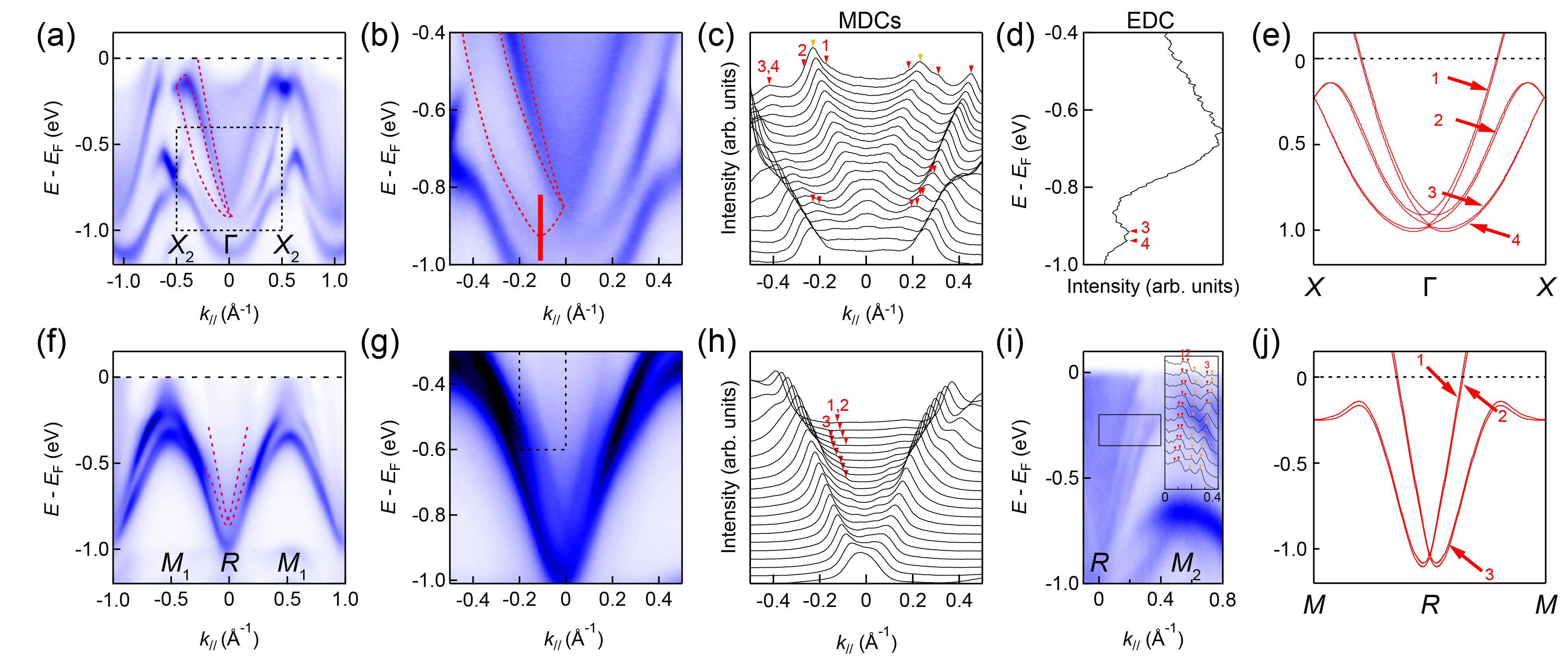}
\caption{
 (a) Band dispersion measured along $X_2$-$\Gamma$-$X_2$ direction (88 eV), red dash lines outline the bulk bands.
 (b) Zoomed in spectra of the dark dashed frame in (a).
 (c) MDCs of (b).
 (d) EDC along the red solid line in (b).
 (e) Bulk bands calculation along the $X$-$\Gamma$-$X$ direction.
 (f) ARPES intensity plot of band dispersion measured along $M_1$-$R$-$M_1$ direction (71 eV), red dash lines outline the bulk bands.
 (g) Zoomed in spectra of the dark dashed frame in (f).
 (h) MDCs of (h).
 (i) ARPES intensity plot of band dispersion measured along $M_2$-$R$-$M_2$ direction (71 eV), inset is the MDCs of the dark frame frame in the same figure.
 (j) Bulk band calculation along the $M$-$R$-$M$ direction.
}
\label{fig:Figure S2}
\end{figure*}

The dispersions along the $M_2$-$R$-$M_2$, $X_1$-$\Gamma$-$X_1$, and $X_2$-$\Gamma$-$X_2$ directions taken at various photon energies are presented in Fig. S3. In Fig. S3(a), the bulk bands indicated by red arrows show significant shape variations across different photon energies, whereas the surface state bands marked by yellow arrows remain largely unchanged. Similarly, in Fig. S3 (c), there is a clear distinction between the bulk bands near 75 eV and those near 88 eV. In Fig. S3(e), the bulk bands near the degeneracy points at 78 eV and 112 eV appear blurred, while those near the $\Gamma$ high-symmetry point at 88 eV remain relatively well-defined.

\begin{figure*}[htb]
\centering
\includegraphics[width=14cm]{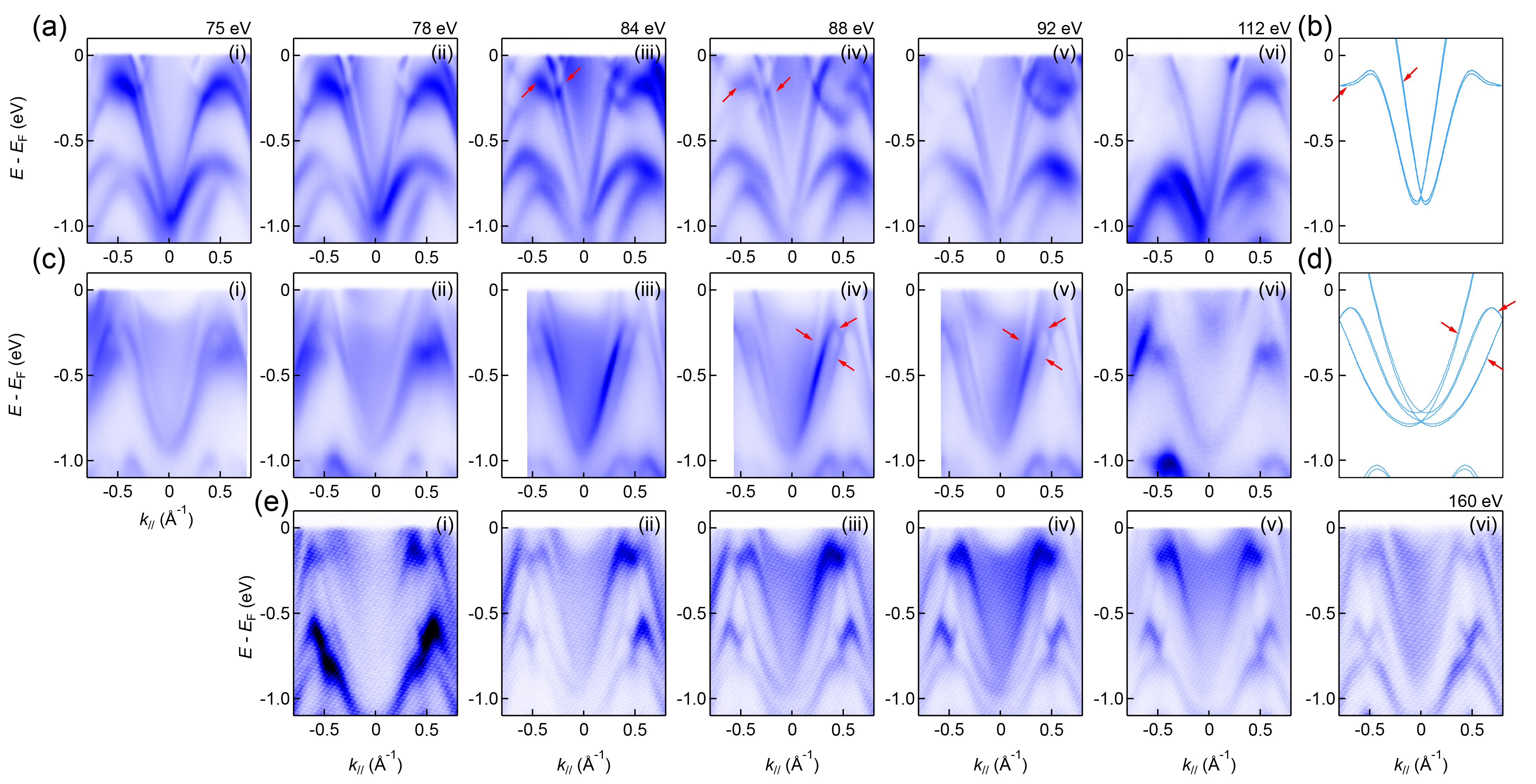}
\caption{
ARPES intensity plot of band dispersion cut along the high symmetry directions with different incident photons. 
(a) ARPES intensity plots of band dispersion cut along the $M_2$-$R$-$M_2$ (projected $\overline{X}_2$-$\overline{M}$-$\overline{X}_2$) direction taken at (i) 75 eV, (ii)78 eV, (iii)84 eV, (iv) 88 eV, (v) 92 eV and (vi) 112 eV.
(b) Bulk band calculations along the $M$-$R$-$M$ direction.
(c) ARPES intensity plots of band dispersion cut along the $X_1$-$\Gamma$-$X_1$ (projected $\overline{X}_1$-$\overline{\Gamma}$-$\overline{X}_1$) direction taken at (i) 75 eV, (ii)78 eV, (iii)84 eV, (iv) 88 eV, (v) 92 eV and (vi) 112 eV.
(d) Bulk band calculations along the $X$-$\Gamma$-$X$ direction.
(e) ARPES intensity plots of band dispersion cut along the $X_2$-$\Gamma$-$X_2$ (projected $\overline{X}_2$-$\overline{\Gamma}$-$\overline{X}_2$) direction taken at (i) 78 eV, (ii)84 eV, (iii)88 eV, (iv) 92 eV, (v) 112 eV and (vi) 160 eV.
}
\label{fig:Figure S3}
\end{figure*}

\section{Surface state bands analysis}

In Figs. S4(a-d), we compare the Fermi surface intensity spectra measured using linear horizontal (LH), linear vertical (LV), right circular (C+), and left circular (C-) polarizations. While the band intensity varies across these polarizations due to photoelectron matrix-element effects, the pronounced chiral feature on the Fermi surface remains consistent, confirming its intrinsic nature.
Furthermore, in Figs. S4(e-f), we compare the constant-energy contour and slice plots measured under LV polarization with those presented in Fig. 3 of the main text. A distinct antisymmetric intensity pattern is observed on either side of the $M$ point, which aligns with the chiral nature of the crystal structure and is independent of the polarization used.

\begin{figure*}[htb]
\centering
\includegraphics[width=14cm]{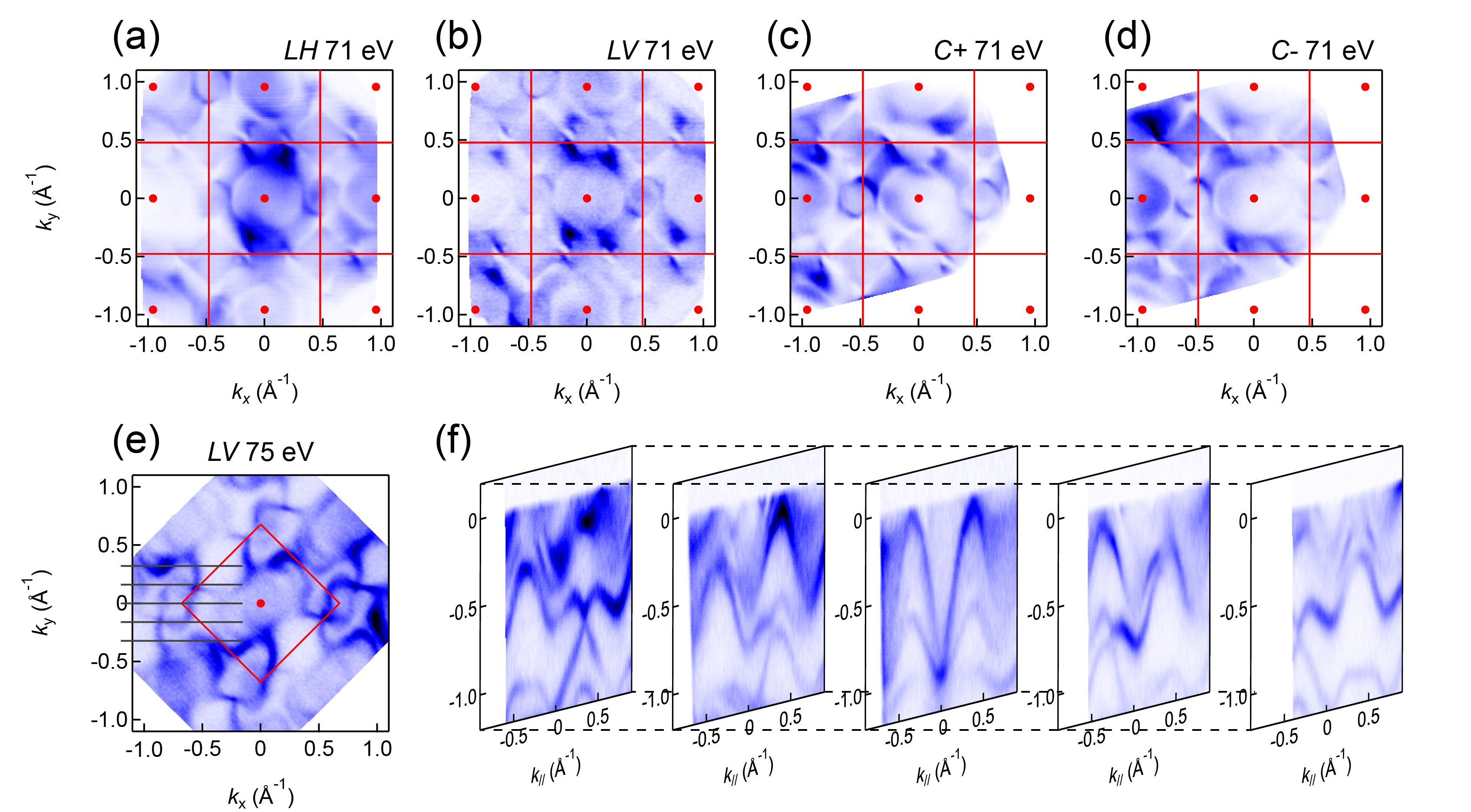}
\caption{
Polarization Analysis. 
(a-d) Fermi surface maps taken by 71 eV photons with linear horizontal (LH), linear vertical (LV), clockwise circular (C+) and anticlockwise circular (C-) light polarization.
(e) Constant energy contour at binding energy of $E_B$ = 0.63 eV taken with 75 eV and LV light polarization.
(f) ARPES intensity plot of band dispersion cuts along the diagonal direction of the Brillouin zone parallel to $\overline{\Gamma}$-$\overline{M}$-$\overline{\Gamma}$ direction.
}
\label{fig:Figure S4}
\end{figure*}

\begin{figure*}[htb]
\centering
\includegraphics[width=14cm]{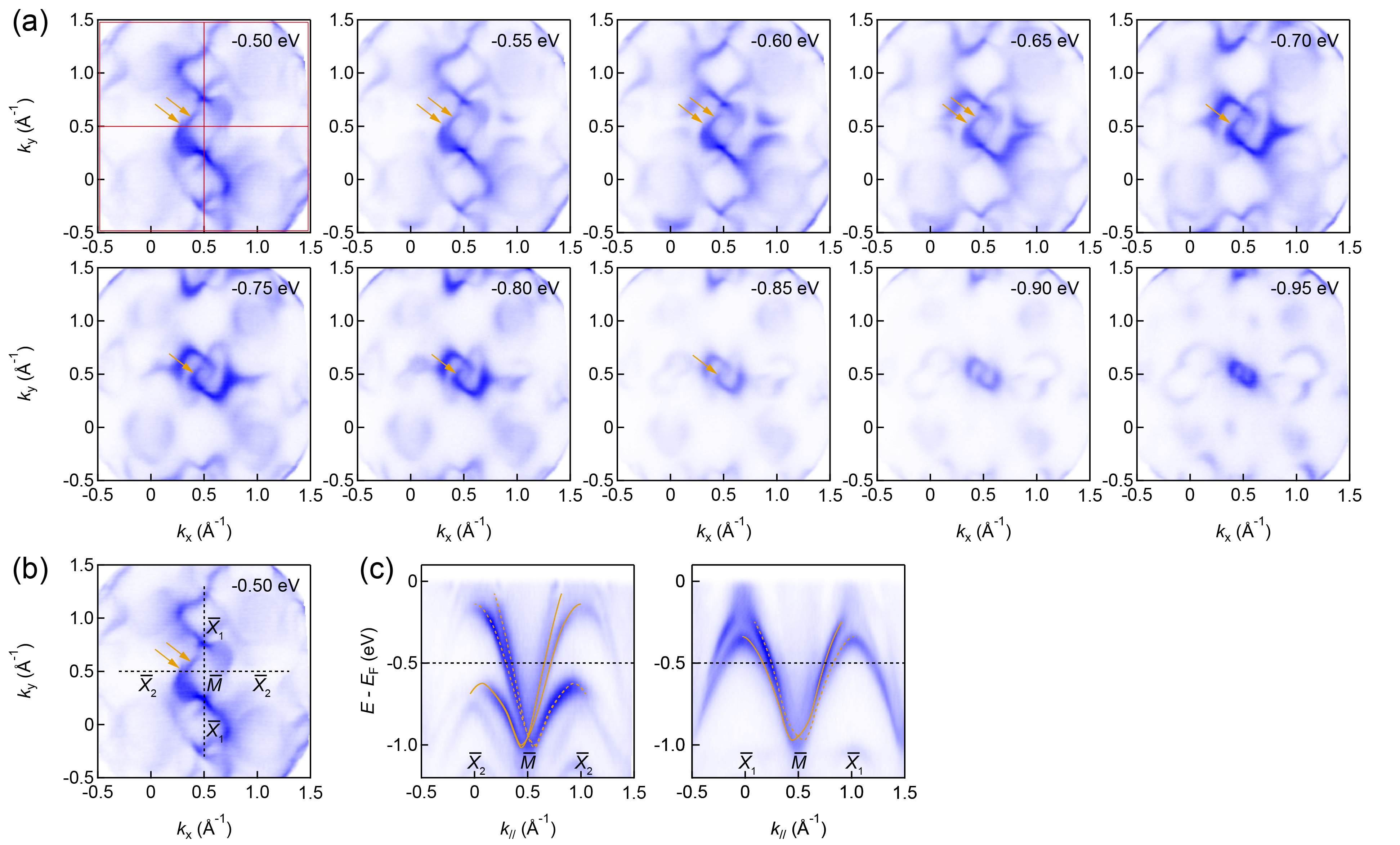}
\caption{
(a) Constant energy counters taken at binding energies from 0.5 to 0.95 eV by step size of 0.05 eV. 
(b) Constant energy counters at $E_B$ = 0.5 eV marked with high symmetry points, yellow arrows indicate the splitting of chiral topological surface states. 
(c) Band dispersions along the $\overline{X_2}-\overline{M}-\overline{X_2}$ and $\overline{X_1}-\overline{M}-\overline{X_1}$ directions, yellow solid and dash lines outline the chiral topological surface states.
}
\label{fig:Figure S5}
\end{figure*}

Depending on the lattice structure of PdSbSe, there may be a variety of cleaving surfaces. We identified three distinct cleaved surfaces, which are labeled as cut-edge 1, 2, and 3 in Fig. S6(a). In Fig. S6(b-e), we compared the band calculations along the $\overline{\Gamma}-\overline{X_2}$ from the experiment with those from theoretical calculations for the three different cleaved surfaces along this direction. It is evident that the position and shape of the surface states vary significantly. However, in all the calculations, only the surface of cut-edge 1 yields a trivial Surface State located below all the bands (indicated by a green dashed line), while the others do not. This explains the bright band observed at -1.0 eV. Similarly, in Fig. S6(f-i), we compared the band structures along the $\overline{M}-\overline{X_1}$. Although the shape and realization of the topological surface states are more similar in some directions, none can account for the surface states near -1.0 eV. Additionally, the position of the surface state bands is influenced by factors such as surface relaxation, surface charge, and the breaking of surface chemical bonds. However, incorporating these factors would significantly increase the computational load. Since the surface states in this system extend throughout the entire Brillouin zone, accurately capturing their positions computationally is indeed challenging.

\begin{figure*}[htb]
\centering
\includegraphics[width=14cm]{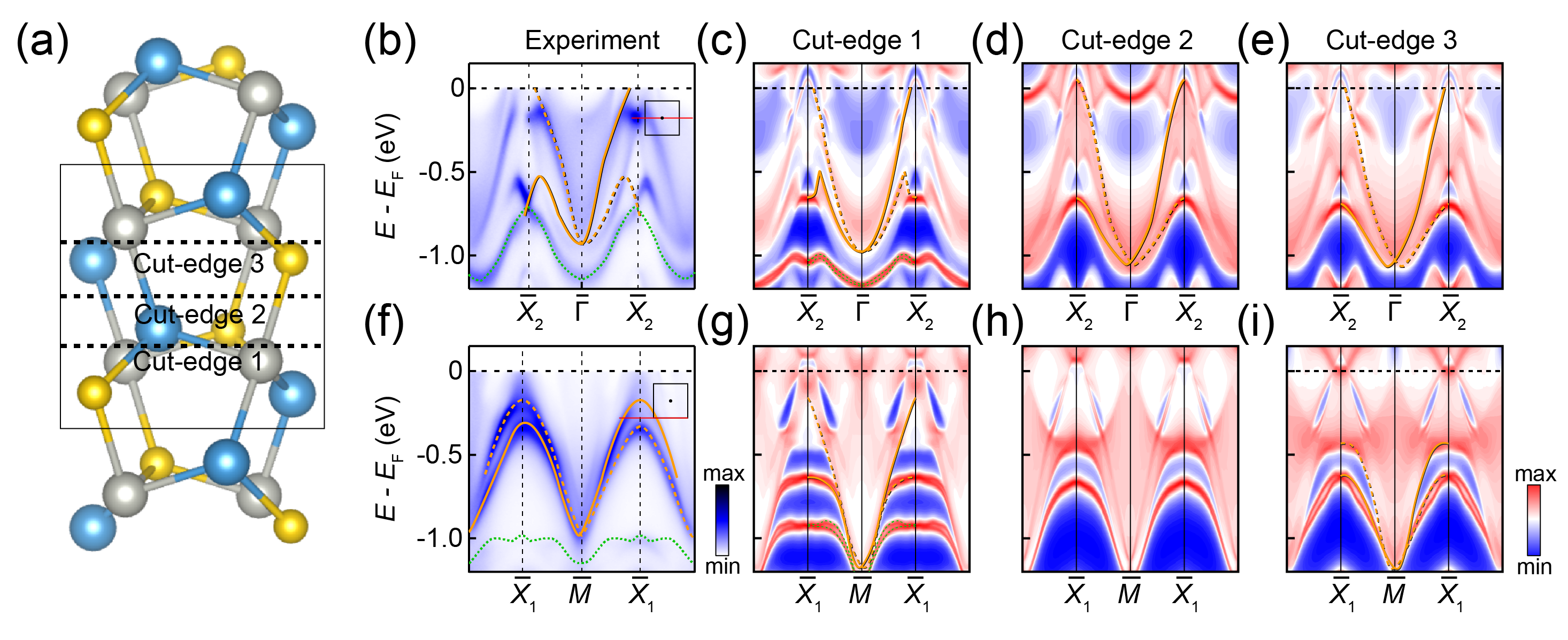}
\caption{
(a)Illustration of different terminations in PdSbSe, labeled by cut-edge 1 to 3. 
(b) Band dispersion along the $\overline{X_2}-\overline{\Gamma}-\overline{X_2}$ direction. 
(c)-(e) Band structure calculation along the $\overline{X_2}-\overline{\Gamma}-\overline{X_2}$ direction on cut-edge 1 to 3. 
(f) Band dispersion along the $\overline{X_1}-\overline{M}-\overline{X_1}$ direction. 
(g-i) Band structure calculation along the $\overline{X_1}-\overline{M}-\overline{X_1}$ direction on cut-edge 1 to 3.
}
\label{fig:Figure S6}
\end{figure*}

\newpage
\section{Domains analysis}
In our analysis of the majority of tested samples, we identified a single chiral domain. However, during certain tests involving different sample batches, we encountered evidence of two distinct chiral orientations, as depicted in Fig. S7(a) and S7(b). These figures illustrate constant-energy surfaces at $E_B$ = 0.63 eV, where petal-shaped patterns with opposite chiral orientations are observed at the Brillouin zone corners. The two patterns exhibit mirror symmetry, which may be attributed to the front and back surfaces of the sample.
We conducted additional experiments to further probe the potential existence of chiral domains. Although we did not directly identify clear chiral domains in most cases, we observed some intriguing phenomena. For instance, in earlier experiments, we obtained perfectly symmetric constant-energy surface data, particularly evident at $E_B$= 0.9 eV, as shown in Fig. S7(d). By comparing this with data from a single chiral cleaved surface (Fig. S7(c)), we surmise that the symmetric data may result from the probe simultaneously detecting signals from two distinct chiral surfaces states. This suggests the possibility of two chiral domains coexisting in the sample, although most of our data primarily reflects a single chiral cleaved surface.

\begin{figure*}[htb]
\centering
\includegraphics[width=8cm]{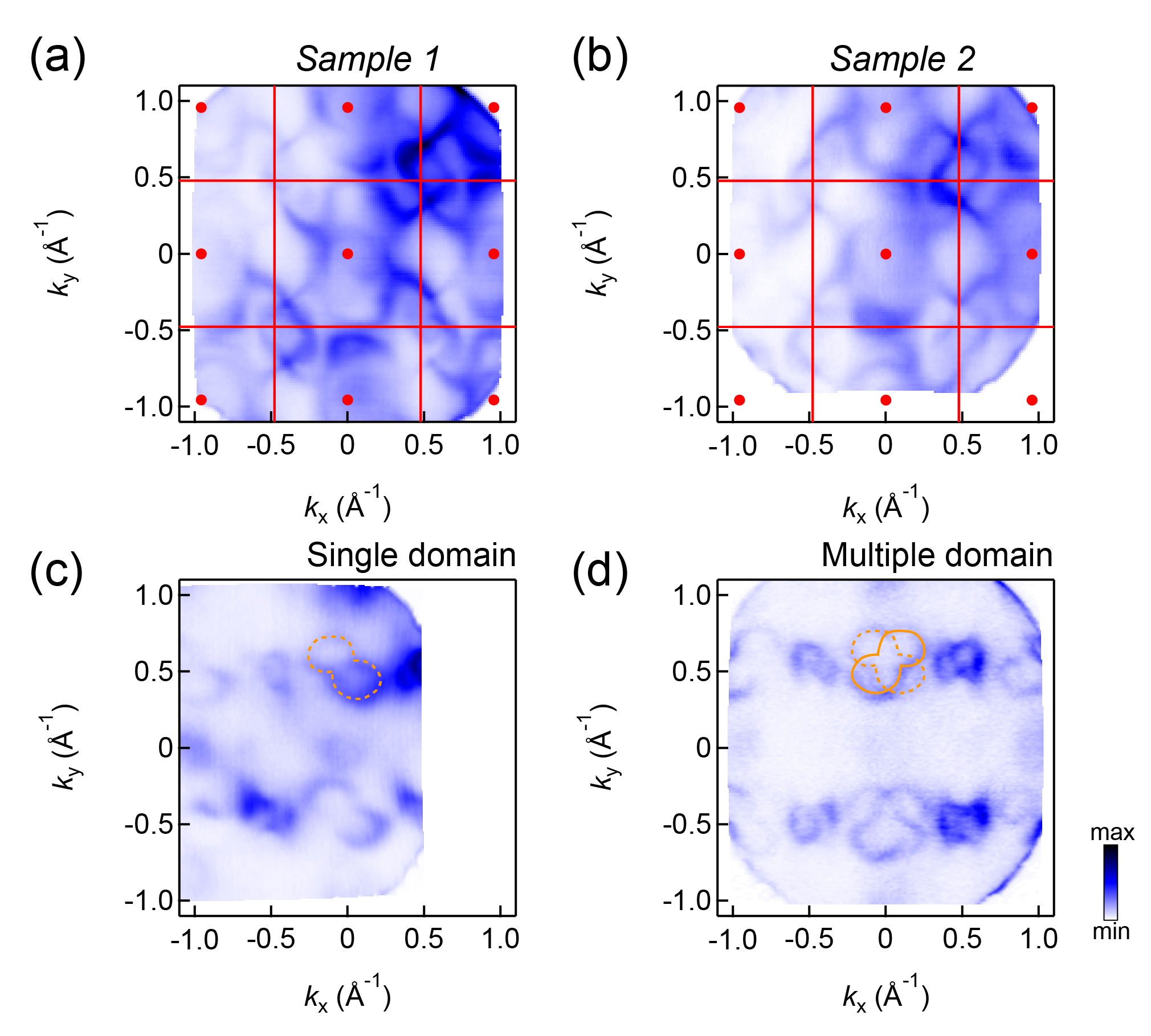}
\caption{
Chiral domains in PdSbSe. 
(a)(b) Constant energy counters at $E_B$ = 0.63 eV taken in sample 1 and sample 2 with different chiral orientation. 
(c) The constant energy counter at EB = 0.9 eV with single chiral domain. 
(d) The constant energy counter at $E_B$ = 0.9 eV with double chiral domain. The data in Fig. S7 were collected by 71 eV photons.
}
\label{fig:Figure S7}
\end{figure*}

\end{document}